
\def\lsim{\mathrel{\scriptstyle{\buildrel < \over \sim}}}
\magnification 1200
\baselineskip=17pt

\centerline{\bf CONFINEMENT OF SPIN AND CHARGE IN}
\bigskip
\centerline{\bf HIGH-TEMPERATURE SUPERCONDUCTORS}
\vskip 50pt
\centerline{J. P. Rodriguez$^{(a)}$ and Pascal Lederer$^{(b)}$}
\smallskip
\centerline{$^{(a)}$\it Theoretical Division,
Los Alamos National Laboratory,
Los Alamos, NM 87545 and}
\centerline{{\it Dept. of Physics and Astronomy,
California State University,
Los Angeles, CA 90032.}
}
\medskip
\centerline{{$^{(b)}$\it Laboratoire de Physique des Solides,
Universit\' e Paris-Sud,
91405 Orsay C\' edex, France.}
\footnote{$^{\dag}$}{Laboratoire associe au CNRS.}}

\vskip 30pt
\centerline  {\bf  Abstract}
\vskip 8pt\noindent
By exploiting the internal gauge-invariance intrinsic
to a spin-charge separated electron,
we show that such  degrees of freedom must be confined
in two-dimensional superconductors experiencing strong
inter-electron repulsion.  We also demonstrate that
incipient confinement in 
the normal state can
prevent chiral spin-fluctuations from destroying
the cross-over
between strange and psuedo-gap regimes
in under-doped high-temperature superconductors.
Last, we suggest that the negative Hall anomaly
observed in these materials is connected with 
this  confinement  effect.
\bigskip
\noindent
PACS Indices:  74.20.De, 74.20.Mn, 74.20.Kk, 71.27.+a
\vfill\eject
The electronic conduction in high-temperature
superconductors is characterized by quasi two dimensionality 
and strong inter-electron repulsion.  Motivated by the corresponding
situation in one dimension, Anderson proposed early on that the 
correlated electron in these materials factorizes into
independent spin and charge degrees of freedom as  a
result.$^1$   The abelian gauge-field theory formulation of the 
$t-J$ model for strongly interacting two-dimensional (2D) electrons
provides an elegant expression of  these ideas.$^{2-4}$
The electron field in such theories is divided up into fermionic
spin and bosonic charge parts, $c_{i\sigma} = f_{i\sigma} b_i^{\dag}$,
where the internal gauge invariance
$(f_{i\sigma}, b_i)\rightarrow e^{i\theta_i}
(f_{i\sigma}, b_i)$ enforces the constraint
$$\sum_{\sigma} f_{i\sigma}^{\dag}f_{i\sigma} + b_i^{\dag}b_i = 1 \eqno (1)$$ 
against
double occupancy at each site. Here, both the spinon field $f_{i\sigma}$ and
the holon field $b_i$ are treated as dynamical quasiparticles
that interact with the {\it statistical} gauge field $a_{\mu}$ associated
with the latter.
Apart from its formal appeal, the gauge-field approach to 
2D electron liquids successfully accounts for many of the 
strange normal-state properties characteristic 
of high-temperature superconductors; e.g., 
the paradoxical observation of a large Luttinger Fermi surface
in conjunction with a hole-type Hall effect, and the unconventional
$T$-linear resistivity.$^4$

Analogous Ginzburg-Landau type gauge-field theories for spin-charge separated
superconductivity in two dimensions
have also been proposed in order to account for the high-$T_c$ phase
diagram (see Fig. 1).$^{5-7}$
Separate superfluid mean-field transitions for the spinon and holon sectors
at respective critical temperatures of $T_f$ and $T_b$ provide the common basis
for these theories.  The Ioffe-Larkin composition formula$^2$
$\rho = \rho_b + \rho_f$ for 
the total resistivity implies a meanfield superconducting 
transition temperature of $T_{c0} = {\rm min} (T_f, T_b)$, while
$T_{*0} = {\rm max} (T_f, T_b)$ marks the cross-over temperature scale
above which strange metallicity (e.g., $\rho\propto T$) sets in.
The latter cross-over phenomenon is consistent with a host
of normal-state properties generally observed in
high-temperature superconductors.$^{8,9}$
It has been recently pointed out,$^{10}$
however, that the entropy generated by the
{\it transverse} component to the statistical gauge-field,
which physically represents chiral spin fluctuations,$^{11}$
completely suppresses the 
spinon-superfluid crossover in
the under-doped regime ($T_b \ll T_f$).  If correct, this
means that a normal-state pseudo-gap$^{12}$ 
is not theoretically possible in  
single-layer oxide superconductors, though one could
still  exist 
in multi-layered compounds.$^{13}$

In general, however,
the longitudinal component of the statistical gauge field must also
be taken into
account in order to properly enforce the constraint (1) against double 
occupancy.  In the case of the
strange-metal saddle-point of the $t-J$ model,  for example,
this component is known to result in 
a negative entropy contribution,$^{14}$ as well as 
a slow-zero sound mode near the
Mott transition.$^{15}$  In the commensurate flux-phase (CFP) 
saddle point,$^{16}$ 
which is believed 
to be an anyonic superconductor,$^{17, 18}$ the longitudinal component leads
to confinement of 
the spinon and the 
holon into the 
electron.$^{19}$
On the basis of a phase-only Ginzburg-Landau
model for spin-charge separated superconductivity
known as the two-component Abelian Higgs (AH$^2$) model,$^7$
and its consistency with the latter,
we shall first show in this paper  
that the spin-charge components of the superfluid
electron are confined in the superconducting phase,
$T < T_c$.  Here,     the superfluid transition at $T_c$
is explicitly of the Kosterlitz-Thouless type.
We shall then demonstrate that incipient confinement
is present in the spin-gap (or Fermi liquid)  regime $T_c < T < T_*$
if   the spinons and  the 
holons couple weakly   to the statistical gauge field.  
This leads to a
drastic  non-perturbative suppression of gauge-field excitations.
(The former can be interpreted as Aslamasov-Larkin type fluctuations$^{20}$
of the confined superconductor within the deconfined normal state.)
We conclude that the spin-gap cross-over at $T_*\lsim T_f$ in the
underdoped regime remains intact for weak-coupling
saddle-points, of which the CFP anyon superconductor
is one.$^{18}$
This means that the 
{\it unique} cross-over
phenomenon observed in a variety of normal-state properties
of under-doped
oxide superconductors$^8$ can in principle be
accounted for by the gauge-field theory
approach to strongly interacting electrons on the square lattice. 
In the case of strong coupling to gauge-field excitations, however, 
we find that
the cross-over into the spin-gap (and Fermi liquid) regimes
is destroyed.$^{7,21}$   This is consistent with recent studies of
the high-temperature strange-metal phase,$^{10}$
which also couples strongly to gauge-field excitations.

{\it Spin-charge Separated Superconductivity.}  The Ginzburg-Landau
energy functional of the AH$^2$ model for spin-charge separated
superconductivity on the square lattice reads$^{7}$
$$\eqalignno{{E} = & J_b\sum_{r,\mu}
\{1-{\rm cos}[\Delta_{\mu}\phi_b-qa_{\mu}]\}+
J_f\sum_{r,\mu}
\{1-{\rm cos}[\Delta_{\mu}\phi_f-qa_{\mu}]\}+\cr
&+\chi_d\sum_{r}
\{1-{\rm cos}[\Delta_xa_y
-\Delta_ya_x]\}, & (2)\cr}$$
where $\phi_b(r)$ and $\phi_f(r)$
represent the respective phases of the
holon and spinon order parameters,
$a_{\mu}(r)=a_{\vec r,\vec r+\hat\mu}$
denotes the statistical
gauge-field, and $\Delta_{\mu}$  denotes
the lattice difference operator ($\mu=x,y$).
The first two terms above are the 
stiffness energies for the phase fluctuations,
where the  local  rigidity of each specie
$i = f, b$ is
related to the critical temperature of its  presumed Kosterlitz-Thouless
(KT) transition  by
$k_B T_i = {\pi\over 2} J_i$.
The last term above  is the stiffness energy for
gauge-field fluctuations,
where the corresponding local
rigidity is given by the sum
$\chi_d = \chi_f + \chi_b$ 
of the diamagnetic susceptibilities of each species,$^{4, 18}$
$\chi_f$ and $\chi_b$.  We assume that the superfluidity
in both the spinon and the holon subsystems
results from Cooper pairing, hence the choice $q=2$.
If we integrate out first the gauge field excitations
from the corresponding partition function
$$Z=\int{\cal D}\phi_b {\cal D}\phi_f {\cal D}a_{\mu}
{\rm exp} (-E/k_B T)\eqno (3)$$
in the continuum limit,
where $E =  {1\over 2}\int d^2r
[J_b(\vec\nabla\phi_b-q\vec a)^2+J_f(\vec\nabla\phi_f-q\vec a)^2]$,
we obtain 
an  effective free energy 
$E_{\rm el} = {1\over 2}\bar J\int d^2r
[\vec\nabla\phi_{\rm el}^{(v)}]^2$ for configurations of the
physical electronic
phase $\phi_{\rm el} = \phi_f - \phi_b$ that carry 
non-zero vorticity ($v$),$^7$ where 
$\bar J = (J_f^{-1}+J_b^{-1})^{-1}$.
This means that the spin and charge of the superfluid electron 
are confined while traversing
the KT transition at
$T_c = (T_f^{-1} + T_b^{-1})^{-1}.$
If, on the other hand, we first integrate out the phase (Higgs)
fields from (3), we then obtain the effective free energy
$$E_{\rm st} = {1\over 2}\mu_{\rm st}^{-1}\chi_d
\int d^2 r [\partial_x a_y  - \partial_y a_x]^2\eqno (4)$$
for the statistical gauge field in the continuum limit, where each
Higgs field contributes a diamagnetic renormalization
$\mu_f^{-1}, \mu_b^{-1} > 0$
to the total renormalization
$$\mu_{\rm st}^{-1} = 1 + \mu_f^{-1} + \mu_b^{-1}\eqno (5)$$
of the statistical magnetic field energy.
The non-perturbative effects mentioned above are included
by making the replacement
$\chi_d\rightarrow\chi_d/\mu_{\rm st}$
in {\it all} perturbative calculations.
Before continuing, let us first identify the statistical London 
penetration length of the AH$^2$ model (2),
$$\lambda_{\rm st} = q^{-1}[\chi_d/(J_f+J_b)]^{1/2}, \eqno (6)$$
here given in units of the lattice constant $a$. 
In the weak-coupling limit $\lambda_{\rm st}
\rightarrow \infty$,
we recover separate superfluid transitions for each species. This 
implies that $\mu_{\rm st} = 0$ at  temperatures 
$T < T_{*0} = {\rm max} (T_f, T_b)$.  
Below, we will demonstrate that 
$T_{*0}$ in fact only marks a sharp cross-over$^{6,9,21}$ for 
finite $\lambda_{\rm st}\gg a$.  It will then be demonstrated that
the CFP anyon superconductor lies inside the latter weak-coupling regime.

Recall that 
if one of the two space     dimensions of the AH$^2$ model is
considered as imaginary time, then
the pure gauge-field 
term in (2) describes vacuum
electromagnetism in one dimension.
Since the latter is trivially confining, the
Wilson loop  for the AH$^2$ model (2) is in general 
related to the
effective diamagnetic renormalization (4) 
due to the Higgs fields  by$^{21}$
$$\Bigl\langle{\rm exp}\Bigl(ip\oint_C
a_{\mu}dx_{\mu}\Bigr)\Bigr\rangle = 
{\rm exp}\Bigl(-{1\over 2}\mu_{\rm st}p^2g^2 A\Bigr) \eqno (7)$$ 
in the limit $p\rightarrow 0$,
where $A$ denotes the area contained by a large
contour $C$ and $g^2 = k_B T/\chi_d$.
Employing the Villain form of the AH$^2$ model (2),$^7$
it can be shown that
the above Wilson loop $W(C)$ is given by the 
average
$$W(C)=\Biggl\langle{\rm exp}
\Biggl[-2\pi i {p\over q}(J_f+J_b)^{-1}
\sum_{r\ {\rm in}\ C}
[J_f q_f(r)+ J_b q_b(r)]\Biggr]\Biggr\rangle_{\rm CG}\eqno (8)$$
over the
corresponding Coulomb gas ensemble
$$Z_{\rm CG}=\sum_{\{q_b, q_f\}}
{\rm exp}\Biggl\{-(2\pi)^2\bar\beta 
\sum_{(r,r^{\prime})}
[G_{\rm lr}(\vec r-\vec r\,^{\prime})
q_{\rm el}(r)q_{\rm el}(r^{\prime})+
G_{\rm sr}(\vec r-\vec r\,^{\prime})
q_{\rm st}(r)q_{\rm st}(r^{\prime})]\Biggr\},
\eqno (9)$$
where the physical electronic
(el) flux-charge and the 
statistical (st) flux-charge are given respectively by
linear combinations
$$\eqalignno{q_{\rm el}=&q_f-q_b,&(10)\cr
     q_{\rm st}=
&\Biggl({J_f\over{J_b}}\Biggr)^{1/2}q_f
+\Biggl({J_b\over{J_f}}\Biggr)^{1/2}q_b,&(11)\cr}$$
of integer fields $q_f(r)$ and $q_b(r)$.
Here $(r,r^{\prime})$
denote combinations
of points covering the dual square lattice,
while $\bar\beta = \bar J/k_B T$.
Corresponding to these charges are long-range (lr)
and short-range (sr) potentials 
 $$\eqalignno{G_{\rm lr}(\vec r)
 =&\int_{\rm BZ}{d^2k\over{(2\pi)^2}}
 e^{i\vec k\cdot\vec r}{1\over{\tilde k^2+\xi_{\rm el}^{-2}}},&(12)\cr
             G_{\rm sr}
 (\vec r)=&\int_{\rm BZ}{d^2k\over{(2\pi)^2}}
 e^{i\vec k\cdot\vec r}
 {1\over{\tilde k^2+\lambda_{\rm st}^{-2}}},&(13)\cr}$$
where 
$\tilde k^2=4-2{\rm cos} \, k_x a -2{\rm cos}\, k_y a$,
and where $\xi_{\rm el}$ denotes the correlation length of the
physical electronic condensate that diverges
in the normal state at $T_c$.  It is known that the Wilson loop
(8) shows a perimeter law in the superconducting phase.$^7$ 
By (7), this implies that $\mu_{\rm st} = 0$ for $T < T_c$;  
i.e., fluctuations of the
statistical magnetic field in spin-charge separated 
superconductors are entirely suppressed!
In the normal state, 
the correlation length $\xi_{\rm el}$ is exponentially large but finite
at  temperatures close to $T_c$. 
Since global vortex charge neutrality is then no longer enforced, the most
important configurations per site are $q_f = 0, \pm 1$, but with
$q_b = q_f$ so that there be no net electronic vorticity $q_{\rm el}$.
Standard manipulations$^{22}$
of expressions (8) and (9) then yield that the diamagnetic renormalization
(7) is given by
$$\mu_{\rm st} (T_c+) = 16\pi\beta_c
\lambda_{\rm st}^{-2(\beta_c -1)},
\eqno (14)$$
where $\beta_c = [(T_b/T_f)^{1/2} +(T_f/T_b)^{1/2}]^2$.
At optimum doping, $T_b = T_f$, we therefore have
${\rm max}\, \mu_{\rm st}\, (T_c+) = 64\pi\lambda_{\rm st}^{-6}$,
which implies a small jump in chiral spin fluctuations
by Eq. (4).$^{19}$ 
Also, Eqs. (4) and (14) 
indicate that chiral spin fluctuations are suppressed   
($\mu_{\rm st}\rightarrow 0$)
exponentially fast   as
the system deviates from the optimum   $T_c$ 
along the superconducting phase boundary.
This is consistent with the narrow window in composition
centered at optimum doping
through which strict $T$-linear resistance is observed
in the normal state of high-temperature superconductors.$^8$

Consider now the spin-gap regime $T_b\ll T_f$ in the same weak-coupling
limit $\lambda_{\rm st}\gg a$. Then the diamagnetic renormalization
due to the spinon subsystem is approximately given by that
of the one component model$^{21-23}$ ($J_b = 0$),
which is known to be 
$1+\mu_f^{-1}  = (16\pi)^{-1}(T/T_f)\lambda_{\rm st}^{2[(T_f/T) - 1]}$.
On the other hand, at temperatures $T\gg T_b$,
a direct high-temperature series analysis of 
the holon term in (2) yields
that$^{23}$ $\mu_b^{-1}\cong {1\over 4}q^2g^2(J_b/k_B T)^4$. After
summing   the latter contributions to obtain 
the total diamagnetic renormalization (5),
the relationship $(\partial\, {\rm ln}\, \mu_{\rm st}/\partial\, {\rm ln}\,
\lambda_{\rm st})|_{T_*} = 0$ then yields 
$T_* \cong T_f[1 - 32\pi^{-2}\lambda_{\rm st}^{-2}(T_b/T_f)^4]$ for the
cross-over temperature, with
$$\mu_{\rm st} \cong 16\pi(T_f/T)\lambda_{\rm st}^{-2[(T_*/T) - 1]}.
\eqno (15)$$
Last, the application of the previously mentioned high-temperature
series results yields that 
$\mu_{\rm st} \cong [1 + {1\over 4}q^2g^2(J_f/k_B T)^4
+ {1\over 4}q^2g^2(J_b/k_B T)^4]^{-1}$ for $T\gg T_{*0}$ in
general.  The overall picture we obtain in the weak-coupling
limit, therefore, is that there is a
sharp cross-over$^{6,9}$ into incipient confinement ($\mu_{\rm st}^{-1}\gg 1$)
as temperature  falls below $T_*$,$^{21}$
followed by true confinement ($\mu_{\rm st} = 0$) 
as temperature falls  below $T_c$. In particular, 
the $T$-linear resistivity contribution due to 
transverse gauge-field fluctuations$^{4}$ (4),
$\rho_b\propto \mu_{\rm st}T/\chi_d$, that is dominant in 
the strange-metal phase is exponentially
suppressed in the spin-gap regime.$^{9, 21}$
Also, the entropy at temperatures just above
$T_{*0}$ due to such transverse gauge field excitations
is approximately$^4$ $S_{\perp}\sim k_B (k_B T_{*0}/\chi_d)^{2/3}
\sim k_B(q\lambda_{\rm st}/a)^{-4/3}$, which is small
in the present weak-coupling limit, $\lambda_{\rm st}\gg a$. 
As a result, the cross-over into the spin-gap (or  Fermi-liquid) regime 
remains intact (see Fig. 1).  This is contrary
to what occurs in the strong-coupling limit,$^{7, 10}$ as discussed   below.

Let us now analyze the 
(high-temperature) strong-coupling limit, $\lambda_{\rm st}\ll a$,
that corresponds to the strange-metal phase,$^{2-4}$
where $\chi_d\propto T^{-1}$.
It is then more convenient to express the Wilson loop (7) in terms of the
original roughening model from which the Coulomb gas ensemble
(9) is derived:$^{7, 21, 22}$ i.e., $W(C) = Z[J]/Z[0]$, with
$$\eqalignno{Z[J] =\sum_{\{n_b, n_f\}}
{\rm exp}\Biggl\{ & -{1\over{2\beta_b}}
\sum_{r,\mu}[\Delta_{\mu}n_b]^2
-{1\over{2\beta_f}}\sum_{r,\mu}
[\Delta_{\mu}n_f]^2\cr
&-{(qg)^2\over 2}\sum_r\Bigl[n_b+n_f+
{p\over	q}J\Bigr]^2
\Biggr\},& (16)\cr}$$
and with 
$J(r) = 0$ unless the
point $r$ lies within the contour $C$,
in which case $J(r) = 1$.  Here, $n_i(r)$
are integer fields and
$\beta_i = J_i/k_B T$.
The present limit
$qg\rightarrow\infty$ requires that
$n_b(r) = -n_f(r)$ for 
$|p/q| < {1\over 2}$.  This  yields
$Z[0] = \sum_{\{n_f\}}
{\rm exp}\{-{1\over 2}\bar\beta^{-1}
\sum_{r,\mu}[\Delta_{\mu}n_f]^2\}$,
as well as an area law (7) for the Wilson loop
with $\mu_{\rm st} =1$.  Hence, while
the cross-over
phenomenon between strange and spin-gap (or Fermi-liquid)
regimes is destroyed,
we  do recover the superfluid transition at $T_c$ from the
former 2D discrete gaussian model (dual to the 2D
$XY$ model).
In other words, the entire normal state
is ``strange'' in the strong-coupling 
limit.$^{7,21}$   This agrees with recent
studies, which begin from the strong-coupling 
strange-metal phase at high temperature,
that     find that
the previously discussed entropy due to transverse
gauge-field fluctuations destroys
the cross-over into the spin-gap regime.$^{10}$  
Note that the evolution from the sharp
cross-over at $T_{*0}$
in weak-coupling to its absence in strong-coupling should be
smooth since the corresponding normal states have a common
high-temperature limit.

{\it Confinement.} 
In summary, the AH$^2$ model for spin-charge separated superconductivity
in two dimensions predicts that chiral spin-fluctuations are
completely suppressed in the superfluid phase.
Such gap-like behavior cannot, however, result from a Higgs mechanism, 
since the only auto-correlation function that
shows algebraic long-range order
in this model
is that corresponding 
to the physical electronic phase
$\phi_{\rm el} = \phi_f - \phi_b$.$^7$  
As a solution to this puzzle, we propose that
the complete suppression of fluctuations in
the statistical magnetic field is a symptom of confinement.$^{17-19}$  
In particular, we suggest that
the dynamics of the statistical
gauge field is described by an  effective action for compact
quantum electrodynamics in two space
dimensions ($x_1, x_2$),
which in the continuum limit reads
$$S_{\rm st} = {1\over {4 g_0^2 a}}\int d^3x (\partial_{\mu} a_{\nu}
-\partial_{\nu} a_{\mu})^2. \eqno (17)$$
Here the time dimension
is rescaled  to $x_0 = c_0 t$ by the velocity of chiral spin-waves,
$c_0$, while $\partial_0 = c_0^{-1}\partial_t$.  Singular 
instanton (monopole)
configurations in (17) act to
confine the statistical electric field
into flux tubes and to suppress
the corresponding magnetic excitations.$^{19}$  However, this occurs only
at temperatures below a   deconfinement temperature,
$T_c > 0$,
that  must necessarily be  identified  with the critical temperature
of the AH$^2$ model (2).
The CFP saddle-point of
the $t-J$ model, which is
considered to be an anyon superconductor,$^{17, 18}$
provides a microscopic realization of this proposal.
The statistical gauge-field is in fact described by
effective action (17) in such case,$^{19}$ with a coupling constant given by
$g_0^{-2} = {2\over\pi} (\Delta_f^{-1} + \Delta_b^{-1})\hbar\omega_0$.
Here $\Delta_b\sim tx$ and $\Delta_f\sim J x^{1/2}$ are    
the respective charge 
and spin gaps for the holon and spinon quasi-particles
at small concentrations, $x$, of mobile holes,$^{18}$
while $\omega_0 = c_0/a$ is the Debye frequency
cut-off of the theory (17).$^{19}$
Employing
the strong-coupling limit 
formula $k_B T_c\cong {2\over{\pi}} \hbar\omega_0 g_0^2$ 
for the deconfinement temperature then yields
$k_B T_c = (\Delta_f^{-1} + \Delta_b^{-1})^{-1}$, which
is of the same form as the $T_c$ formula for the AH$^2$
model (2).  Also, given 
that $g^{-2} = g_0^{-2}(\hbar\omega_0/k_B T)$
by quantum statistical mechanics, 
we obtain the physically
satisfying result $\lambda_{\rm st}\sim x^{-1/2}$ 
for the statistical length scale (6).  Note that
we have here used the result
$(\hbar\omega_0)^2\sim tJ x^{1/2}$ valid for the CFP near half
filling, and  have 
identified  $k_B T_{i}$ with  $\Delta_{i}$.
Hence, we find that the CFP is in the weak-coupling
regime, $\lambda_{\rm st}\gg a$, near half filling,
$x\ll 1$.

In conclusion, overall consistency of the AH$^2$
model (2) requires that the
spin and charge parts of the superfluid electron
be confined in two dimensions.  This indicates
that the superfluid component, $\sigma_{xy}^s$,
of the in-plane Hall conductance in spin-charge
separated superconductors should have an electron
sign.  A negative sign anomaly for $\sigma_{xy}$
is in fact observed in high-temperature
superconductors [$\sigma_{xy}$ exhibits a (positive) hole 
sign in the normal state].  Our result therefore justifies	
the two-fluid explanation for such observations,$^{24}$
wherein the Hall conductance is given by the sum,
$\sigma_{xy} = \sigma_{xy}^{n} + \sigma_{xy}^{s}$,
of a hole-type normal conductance, $\sigma_{xy}^{n} > 0$,
and an electron-type superfluid conductance, $\sigma_{xy}^{s} < 0$.
This proposal of course fails to explain the sign changes
of the Hall effect observed in certain      
low-$T_c$ superconductors.$^{25}$

J.P.R. acknowledges R. Laughlin  and M. Salkola for
valuable discussions. 
He is also grateful for the
hospitality shown by the Laboratoire
de Physique des Solides, Universit\' e de Paris,
Orsay, where part of this work was completed.
This work was performed under
the auspices of the Department of Energy and was supported in
part by NATO grant CRG-920088 and by National
Science Foundation grant DMR-9322427.

\vfill\eject
\centerline{\bf References}
\vskip 16 pt

\item {1.}  P.W. Anderson, Science
 {\bf 235}, 1196 (1987); in ``Frontiers and Borderlines in
Many-body Physics'', Varenna Lectures (North Holland,
Amsterdam, 1987).

\item {2.} L.B. Ioffe and A.I. Larkin,
 Phys. Rev. B {\bf 39}, 8988 (1989).

\item {3.} J.P. Rodriguez and
 B. Dou{\c c}ot, Europhys. Lett.
{\bf 11}, 451 (1990).

\item {4.} N. Nagaosa and P.A.Lee,
Phys. Rev. Lett. {\bf 64},
2450, (1990);  L.B. Ioffe and
 P.B. Wiegmann, Phys. Rev. Lett. {\bf 65},
653 (1990); L.B. Ioffe and
 G. Kotliar, Phys. Rev. B {\bf 42}, 10348 (1990).

\item {5.} S. Sachdev, Phys. Rev.  B {\bf 45}, 389 (1992).

\item {6.} N. Nagaosa and P.A. Lee,
 Phys. Rev. B {\bf 45}, 966 (1992).
 
\item {7.} J.P. Rodriguez, Phys. Rev. B {\bf 49},
3663 (1994); 9831 (1994) [There is an inconsistency in the
latter reference that can be corrected by replacing 
$e^{\prime}$ with $e^{\prime}/g$ in Eq. (46) and within the
text of section V.].	

\item {8.} B. Batlogg {\it et al.}, Physica C {\bf 235}, 130 (1994) and
references therein.

\item {9.} M. Gabay and P. Lederer, Phys. Rev. B {\bf 47}, 14462 (1993).

\item {10.} M.U. Ubbens and P.A. Lee, Phys. Rev. B {\bf 49}, 6853 (1994).

\item {11.} X.G. Wen, F. Wilczek, and A. Zee, Phys. Rev. B {\bf 39},
11413 (1989).

\item {12.} A. Sokol and P. Pines, Phys. Rev. Lett. {\bf 71}, 2813 (1993).

\item {13.} A.J. Millis and H. Monien, Phys. Rev. Lett. {\bf 70},
2810 (1993); {\bf 71}, 210 (E) (1993).

\item {14.} R. Hlubina, W.O. Putikka, T.M. Rice, and D.V. Khveshchenko, 
Phys. Rev. B{\bf 46}, 11224 (1992).

\item {15.}  J.P. Rodriguez, Phys. Rev. B {\bf 44}, 9582 (1991); {\bf 45}, 
5119 (E) (1992);  Phys. Rev. Lett. {\bf 73}, 1675 (1994).

\item {16.} P. Lederer, D. Poilblanc, T.M. Rice,
Phys. Rev. Lett. {\bf 63}, 1519 (1989).

\item {17.} R.B. Laughlin, Science {\bf 242}, 525 (1988); 
in {\it Modern Perspectives in Many-Body Physics}, 
edited by M.P. Das and J. Mahanty 
(World Scientific, Singapore,1994) p. 249. 

\item {18.}  J.P. Rodriguez and B. Dou\c cot,
Phys. Rev. B{\bf 45}, 971 (1992).

\item {19.} J.P. Rodriguez, Phys. Rev. B {\bf 51}, 9348 (1995).

\item {20.} M. Tinkham, {\it Introduction to Superconductivity}
(McGraw Hill, Malabar, 1975), Ch. 7.

\item {21.} J.P. Rodriguez and P. Lederer, Phys. Rev. B {\bf 48},
16051 (1993).

\item {22.} D.R.T. Jones, J. Kogut,
D.K. Sinclair, Phys. Rev. D {\bf 19},
1882 (1979).

\item {23.}  E. Fradkin and S.H. Shenker, Phys. Rev.
D {\bf 19}, 3682 (1979).
 
\item {24.} V.B. Geshkenbein and A.I. Larkin, Phys. Rev. Lett.
{\bf 73}, 609 (1994), and references therein.

\item {25.} See for example S.J. Hagen et al., Phys. Rev. B {\bf 47},
1064 (1993).

\vfill\eject
\centerline{\bf Figure Caption}
\vskip 20pt
\item {Fig. 1.}  Shown is the phase diagram of 2D
spin-charge separated superconductors in the weak-coupling limit,
$\lambda_{\rm st}\gg a$.  The
line that encloses  the superconducting phase represents a
true phase boundary, while the line above it
marks  the 
cross-over point $T_*$.  The monotonically
increasing and decreasing dashed lines represent 
critical temperatures
$T_b = (200\,{\rm K}) x/x_0$ and 
$T_f = (800\,{\rm K})(1 - x/x_0)$, respectively, 
for the holon and spinon subsystems
in isolation, where $x_0$ denotes the mobile hole concentration
below which spinon pairing occurs.

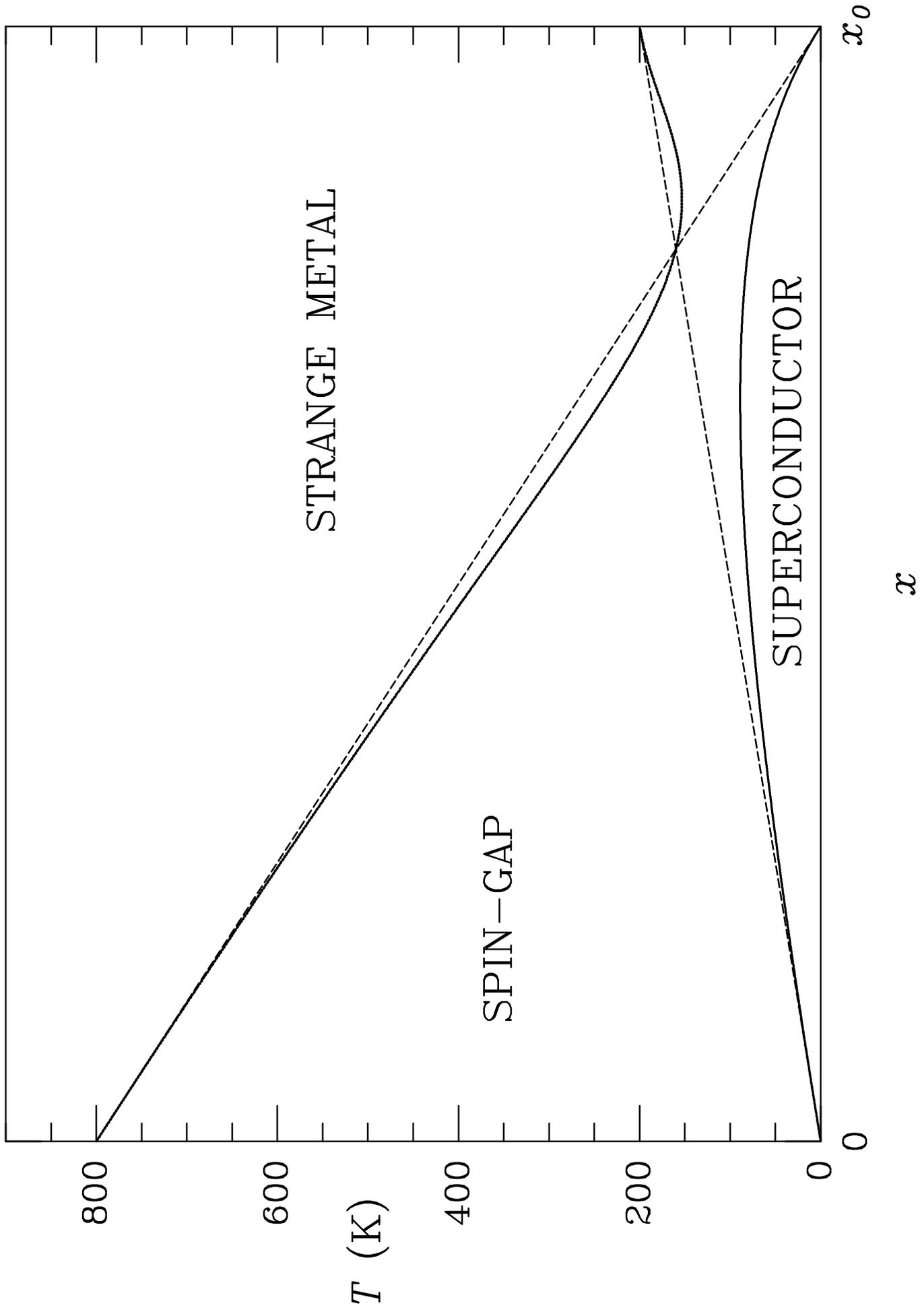
\end